# High quality electron bunch generation with $CO_2$-laser plasma accelerator

L. G. Zhang, B. F. Shen,[*] J. C. Xu,[†] L. L. Ji, X. M. Zhang, W. P. Wang, X. Y. Zhao,

L. Q. Yi, Y. H. Yu, Y. Shi, T. J. Xu, and Z. Z. Xu

*State Key Laboratory of High Field Laser Physics, Shanghai Institute of Optics and Fine Mechanics, Chinese Academy of Sciences, P. O. Box 800-211, Shanghai 201800, China*

$CO_2$ laser-driven electron acceleration is demonstrated with particle-in-cell simulation in low-density plasma. An intense $CO_2$ laser pulse with long wavelength excites wakefield. The bubble behind it has a broad space to sustain a large amount of electrons before reaching its charge saturation limit. A transversely propagating inject pulse is used to induce and control the ambient electron injection. The accelerated electron bunch with total charge up to 10 nC and the average charge per energy interval of more than 0.6 nC/MeV are obtained. Plasma-based electron acceleration driven by intense $CO_2$ laser provides a new potential way to generate high-charge electron bunch with low energy spread, which has broad applications, especially for X-ray generation by table-top FEL and bremsstrahlung.

Laser-driven plasma-based accelerators (LPAs), originally proposed by Tajima and Dawson [1] in 1979, produce ultra-high acceleration gradients (>100 GV/m) for electrons and are promising to make compact accelerators. Over the last three decades, new technologies, especially the chirped-pulse amplification (CPA) method [2] helps to increase the laser intensity dramatically, the LPAs have made tremendous progress in generating high-energy and quasi-monoenergetic electron beams. Various methods were used to generate high-energy electron beams of 1-2 GeV [3-6]. Recently, the electron energy has been boosted over 3 GeV with a single PW laser pulse by using a dual-stage laser-wakefield acceleration scheme [7]. In addition to the electron energy, the total charge and charge per energy interval of the electron bunches are also important for future table-top sources of ultra-short, coherent hard X-rays [8,9] for physical, chemical and biological applications.

Among methods of generating X-rays, bremsstrahlung radiation and X-ray free electron lasers (X-ray FELs) are two effective ways. In the bremsstrahlung, electron bunches incident into and slow down in a conversion target (the "radiator") with high atomic number; to generate X-ray radiation efficiently, the incident bunch containing high charge is required, for example, up to $10^{10}$ (~nC) electrons per shot [10]. Other properties, such as mono-chromaticity, small divergence, pointing stability, etc., on the electron bunches are requested at a moderate level [11]. In the X-ray FEL, in which relativistic electrons undergo transverse oscillations as they pass through the undulator under the Lorentz force of the periodic magnetic field and emit polarized radiation [12], high charge is also demanded. More rigorously, to exceed the threshold for FEL action at X-ray photon energies, the bunches request high qualities with short bunch duration and small energy spread [13], thus with large average charge per energy interval.

Recently, an average charge per energy interval of more than 10 pC/MeV electron beams has been produced experimentally with a shock-front injector by using a 800 nm Ti:sapphire laser [14]. Driven by short-wavelength (e.g. 800 nm) laser pulse, further getting electron beams with more average charge per energy interval, namely, obtaining higher bunch charge and improving its energy spread simultaneously is very difficult because of the limitation of the small bubble scale. To this end, a scheme of laser wakefield accelerator (LWFA) driven by a $CO_2$ laser is proposed in this letter. Nowadays, high-pressure $CO_2$ laser has approached a Terawatt power [15] and has been applied successfully for driving particle acceleration [16-18]. The $CO_2$ laser pulse, with a central wavelength of 10.6 μm, can drive a large-size bubble in low-density plasma, and the bigger bubble volume can accelerate more electrons to high energy before overloading effect occurs. Thus the energy spread and the charge per energy interval will be improved significantly.

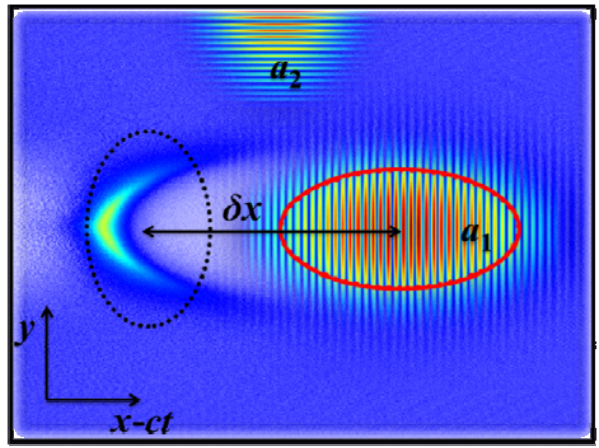

FIG. 1 (color online).   Schematic diagram of the PIC simulation. The bubble is excited by the $CO_2$ drive laser pulse ($\lambda$ = 10.6 μm), propagating along $x$ direction. The transversal inject pulse, usually with a much lower intensity than the drive pulse, incidents into the plasma along $y$ direction. When its peak arrives $y$=0 (the black dotted ellipse), the drive pulse (the red solid ellipse) moves in ahead with $\delta x$ = 6$\lambda$.

This mechanism is explored by using the two dimensional PIC code VORPAL [19] in Cartesian coordinates with a moving window, as depicted in Fig. 1. A relativistic intense drive pulse from $CO_2$ laser with the wavelength $\lambda$ = 10.6 μm and the dimensionless amplitude $a_1$ = 2.0, propagates along $x$ axis, and excites a bubble region moving forward with a phase velocity near the speed of light $c$ in the uniform plasma whose electron density is $n_e$ = 3.42×10$^{16}$ cm$^{-3}$; the electron self-injection does not occur in such a plasma density. The inject pulse moving transversely across the wakefield along the *y* direction is introduced to push the ambient electrons into the bubble. This transversal injection ensures the electrons be injected and begin to be sped up from a certain position and could also avoid some propagation effects [20] compared with the injection method of head on colliding [21,22]. In our simulation, the dimensionless amplitude of the inject pulse $a_2 = 2.0$, equaling to that of the drive pulse, because the inject pulse has to be strong enough that its ponderomotive force can expel quite a few of electrons from the ambient plasma and the bubble sheath into the bubble inside. Both the drive pulse and the inject pulse have a pulse duration $L = 11.2\lambda$ (~0.4 ps) and focus size $w = 15\lambda$ (~160 μm) full-width at half-maximum. Furthermore, the inject pulse is incident into the underdense plasma at $x = 61.5\lambda$ (~652 μm); when it arrives at the axis ($y = 0$), the spatial distance between the center of the inject pulse and the drive pulse is $\delta x = 6\lambda$. This spatial and time delay between the pulses is selected to ensure the effectiveness of injection. The simulation window has a dimension of $60\lambda \times 70\lambda$ with the same single grid size of $\lambda/25$ in the *x* and *y* directions, respectively. The uniform plasma is distributed from $x = 20\lambda$ to $600\lambda$ and $y = -30\lambda$ to $30\lambda$.

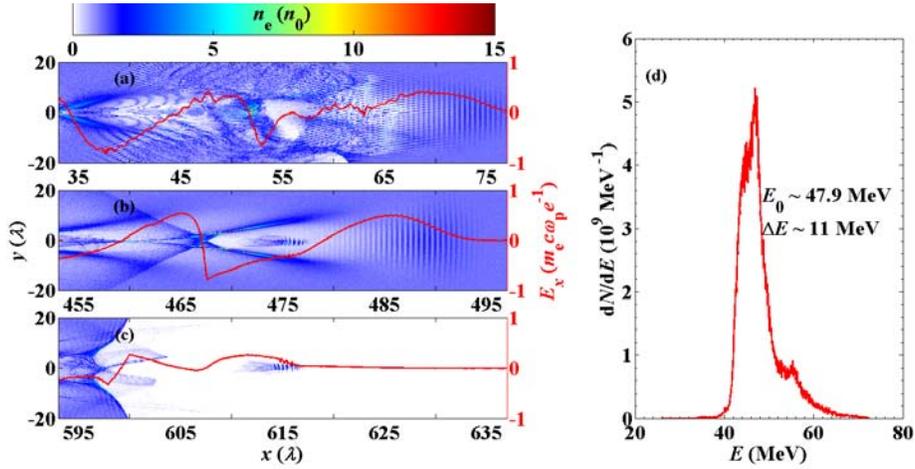

FIG. 2 (color online). Snapshots (a)-(c) show the electron density distribution of the plasma and the acceleration electric field $E_x$ curve along the symmetry axis ($y=0$) of the bubble when the drive pulse centered at different positions. (a) at ~ $70\lambda$, the inject pulse just arrive at $y = 0$, (b) at ~$490\lambda$, an electron bunch has been captured and accelerated in the first bubble region, (c) at ~ $630\lambda$, the bunch just rush out from the plasma and enter the vacuum. Electrons density is normalized to ambient plasma density $n_0$ and the electric field is normalized to the wave breaking field $m_e c \omega_p e^{-1}$.
(d) the energy spectrum of the electron bunch in the first bubble region in Fig. 2(c), the weighted mean energy of the bunch is about 47.9 MeV and the FWHM energy spread is about 11 MeV. Simulation parameters: $a_1 = a_2 = 2.0$ and $\lambda = 10.6$ μm.

Figure 2 illustrates the simulation results with parameters $a_1 = a_2 = 2.0$ and $\lambda = 10.6$ μm. When the inject pulse approaches the symmetry axis ($y = 0$), it has pushed quite a big amount of electrons into the first bubble. Thus, the bubble shape is destroyed as shown in Fig. 2(a), and the electric field $E_x$ has a remarkable increase in the region $x \in (48\text{-}60)$ μm. After the inject pulse travels past the bubble region, the injection ceases. A small part of the electrons pushed by the inject pulse can catch the bubble and gain energy inside continuously. An electron bunch accelerated in the first bubble can be seen clearly in the Fig. 2(b). At the position of the bunch, the $E_x$ curve increases around $x = 475$ μm; here the "beam loading" effect appears and raises the local electric field. Therefore, all electrons in the electron bunch feel a relative same acceleration field and the bunch maintain a relative low energy spread. After the bunch rush out of the plasma [Fig. 2(c)], statistical analysis shows that there are about $3.58 \times 10^{10}$ electrons in the bunch and the total charge $Q$ is about 5.7 nC, the weighted mean energy of the bunch is 47.9 MeV and the full width at half maximum (FWHM) energy spread $\Delta E$ is about 11 MeV. Thus the average charge per energy interval $Q/\Delta E$ is 0.52 nC/MeV, which is much larger than that obtained with 800 nm laser pulses [14]. In addition, from Fig. 2(b), one can see the diameter of the bubble is about 200 μm, that is why it can carry so many electrons with a charge of around 5.7 nC. Meanwhile the electron bunch has a large sizes of ~60 and ~20 μm in $x$ and $y$ directions, respectively, which is easier to be detected and manipulated in the bremsstrahlung applications.

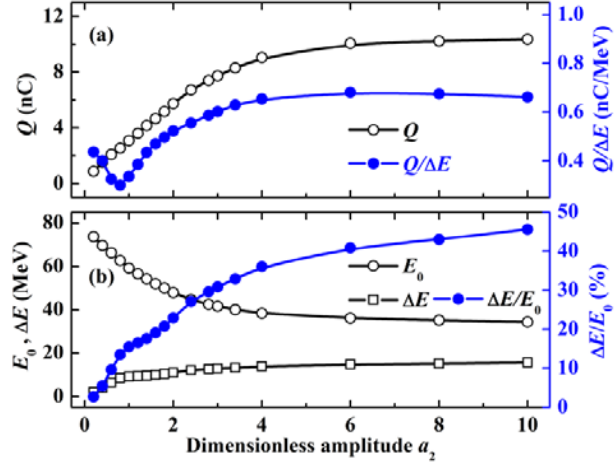

FIG. 3 (color online). (a) Total charge $Q$, average charge per energy interval $Q/\Delta E$, and (b) center energy $E_0$, FWHM energy spread $\Delta E$ and relative energy spread $\Delta E/E_0$ of the electron bunch for various value of the inject pulse $a_2$ when the drive pulse center arrives at ~$630\lambda$ ($\lambda$ = 10.6 μm, $a_1$ = 2.0), the bunch just rushes into the vacuum.

From the inject process mentioned above, the intensity of the inject pulse is an important factor which determines how many electrons can be pushed into the bubble and then influences the energy spread of the accelerated electron bunch. Serious of simulations with different $a_2$ from 0.2 to 10.0 have been performed while keeping other simulation parameters as same as the case $a_2$= 2.0. After the electron bunch rushing out of the plasma at $x = 615\lambda$ [see Fig. 2(c)], main physical quantities of the bunch versus $a_2$ are shown in Fig. 3.

One can find in Fig. 3(a) that the total charge $Q$ of the bunch increases almost linearly when $a_2 < 2.0$, then the growth slows down and finally reaches a saturation value of about 10 nC at $a_2 = 10.0$. When $a_2 = 0.2$, the average charge per energy interval $Q/\Delta E$ is about 0.44 nC/MeV; then it reduces and down to the minimum value of about 0.3 nC/MeV at $a_2 = 0.8$, after that the $Q/\Delta E$ rebounds and reaches about 0.68 nC/MeV at $a_2 = 6.0$; when $a_2 > 6.0$, $Q/\Delta E$ slides down slightly. The variations of the $Q/\Delta E$ curve are a consequence of the total charge $Q$ [in Fig. 3(a)] and the FWHM energy spread $\Delta E$ plotted in Fig. 3(b). One can find that $\Delta E$ increases faster than the total charge $Q$ at first, which corresponding to the first reduction stage of $Q/\Delta E$; then $\Delta E$ remains a gentle rise, therefore, the curve shape of the $Q/\Delta E$ is similar with the total charge when $0.8 < a_2 < 6.0$. The slight reduction of $Q/\Delta E$ when $a_2 > 6.0$ is the result of the well saturation of $Q$ and the continuous increase of $\Delta E$. At the same time, in Fig. 3(b), the mean energy $E_0$ of the electron beam decreases due to the "beam loading" effect and the relative energy spread $\Delta E/E_0$ goes upwards obviously. When $a_2$ reach 10.0, the relative energy spread ascends above 45% and the electron energy reduces to 34.3 MeV.

The bubble excited by $CO_2$ laser pulse has a good potential to sustain and accelerate electron bunch with large amount of charge up to 10 nC. At appropriate intensities of the inject pulse, electron bunches with average charge per energy interval well above 0.6 nC/MeV could be obtained, which is scores of times larger than that obtained with 800 nm laser pulse in Ref. [14].

In order to understand the effect of the tenfold increase of the laser wavelength, we consider a simple model of spherical bubble immersed in the underdense plasma. It is known that a single, short and high intensity laser pulse, with a wavelength $\lambda_i$, drives a plasma wave. The wakefield is driven most efficiently when the laser pulse length is on the order of the plasma period, namely $L_i \sim \lambda_{pi}$, the corresponding density of plasma ambient electrons can be written as

$$n_{ei} = \pi m_e c^2 / (e\lambda_{pi})^2. \tag{1}$$

During the interaction of laser pulse and the plasma, we suppose the ponderomotive force of the laser pulse expels violently all electrons out of the focal spot, leaving in its wake a solitary ion cavity (plasma bubble) around the laser pulse. The bubble shape is determined by ponderomotive potential and the force of the electron bunch accelerated in it. For simplicity, the bubble can be seen as a sphere, whose diameter is on the order of $\lambda_{pi}$ and volume can be estimated by

$$V_i = \tfrac{4}{3}\pi(\lambda_{pi}/2)^3. \tag{2}$$

The number of the electrons expelled by the laser pulse, or the maximum electron capacity of the bubble is

$$N_i = n_{ei}V_i = m_e \pi^2 c^2 \lambda_{pi}/6e^2. \tag{3}$$

Actually, the number of electrons being trapped is just a fraction of the bubble capacity. Supposing the factor of the accelerated electrons over the bubble capacity is $\kappa$, then the total charge of the accelerated electrons is

$$Q_i = e\kappa N_i = m_e \pi^2 c^2 \kappa \lambda_{pi}/6e. \tag{4}$$

From Eq. (4), it is clear that the total charge is proportional to the plasma wavelength, and also to the laser wavelength, provided that $\lambda_{p1}/\lambda_{p2} = L_1/L_2 = \lambda_1/\lambda_2$. Furthermore, only when the accelerated electrons have high charge while retaining a low energy spread, it can achieve a high average charge per energy interval.

The theoretical analysis suggests that there may be a linear relation between the bunch charge and the drive wavelength. Further simulations with drive wavelengths $\lambda_i$ from 0.8 to 12.0 μm have been performed to confirm this relation. The parameters used in these simulations are normalized to the drive wavelength $\lambda_i$, i.e., laser pulse duration $L_i = 11.2\lambda_i$, focus size $w_i = 15\lambda_i$ and electron density of plasma $n_{ei}(cm^{-3}) = 3.84 \times 10^{18}[\lambda_i(\mu m)]^{-2}$, etc. For example, in the case of using a Ti:sapphire laser ($\lambda_i$ = 0.8 μm) as the drive pulse, the plasma with electron density of $6\times10^{18}$ cm$^{-3}$, drive laser pulse with duration of 30 fs and focal spot size of 12 μm (FWHM), and inject pulse incidents at about 50 μm are used. In these simulations, the dimensionless amplitude of the drive and the inject pulse is $a_1 = 2.0$ and $a_2 = 0.2$, respectively.

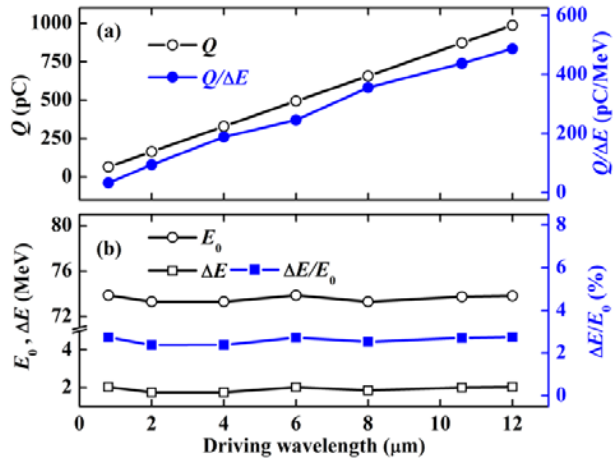

FIG. 4 (color online).  (a) Total charge $Q$, average charge per energy interval $Q/\Delta E$, and (b) center energy $E_0$, FWHM energy spread $\Delta E$ and relative energy spread $\Delta E/E_0$ of the electron bunch versus the drive wavelength $\lambda_i$ when the center of the drive pulse arrives at $\sim630\lambda_i$ ($a_1 = 2.0$,

$a_2 = 0.2$), the bunch just enter the vacuum at this moment.

Physical quantities of the electron bunch discussed above such as $Q$, $E_0$ and $\Delta E$ are displayed in Fig. 4 versus the drive wavelength $\lambda_i$. From Fig. 4(a) one can clearly find that the total charge points align themselves, the correlation coefficient of the total charge and the drive wavelength is very close to 1.0, which indicates that there indeed exists the linear relation as shown in Eq. (4). According to Fig. 4(b), the weighted mean energy $E_0$, the FWHM and relative energy spread of the bunches are almost the same at each drive wavelength, so the average charge per energy interval of the electron bunches present a near-linear relation, with a correlation coefficient of 0.9973, is shown in Fig. 4(a). This relationship indicates a potential method of generating high charge and high quality bunches. The longer the wavelength of the drive pulse is, the higher charge of the bunch will be obtained, meanwhile the energy spread keeps in a low level, provided the intensity of the drive pulse have reached the relativistic intensity. Of course, the wavelength should not be as long as used in conventional accelerators which are big facilities.

In conclusion, the scheme of producing high quality electron bunch with high total charge and high average charge per energy interval in the plasma bubble driven by long wavelength $CO_2$ laser pulse is presented, and the total charge, energy spread and average charge per energy interval of the bunch are studied. Using a transversal propagating inject pulse, electrons injected into the bubble can be controlled by adjusting the dimensionless amplitude of the inject pulse. At $a_1 = a_2 = 2.0$, an electron bunch with total charge $Q \sim 5.7$ nC, $\Delta E \sim 11$ MeV and $Q/\Delta E \sim 0.52$ nC/MeV are obtained. The large-size bubble excited by $CO_2$ laser pulse has a potential in capturing and accelerating large amount of charge. Nearing the charge saturation limit of the bubble, electron bunch carries charge up to ~10 nC and reaches the average charge per energy interval of more than 0.6 nC/MeV. The linear relationship between the total charge and the drive wavelength, verified through a sequence of confirmatory simulations, gives an effective method to get higher quality electron beam by using longer wavelength drive pulse.

This work has been supported by the Ministry of Science and Technology (2011CB808104) and National Natural Science Foundation of China (Projects Nos. 11125526, 11335013, 11374319, 11374317, 11127901, and 61221064). Authors thank Shanghai Supercomputer Center for the support of calculation work.

[*] bfshen@mail.shcnc.ac.cn

[†] jcxu@siom.ac.cn